\documentclass[prd,showpacs,nofootinbib,preprintnumbers]{revtex4}

\usepackage{latexsym}
\usepackage{amssymb}
\usepackage{amsmath}
\usepackage{bm}
\usepackage{float}
\usepackage{indentfirst}
\usepackage{longtable}
\usepackage[dvips]{epsfig}
\usepackage{amscd}

\usepackage{xspace}

\renewcommand{\cite}{\citep}
\newcommand{\comment}[1]{}

\newcommand{\hsp}{\hspace{0.15em}}
\newcommand{\nhsp}{\hspace{-0.15em}}
\newcommand{\smhsp}{\hspace{0.06em}}

\newcommand{\ds}{\displaystyle}

\newcommand{\la}{\lambda}

\newcommand{\om}{\omega}
\newcommand{\al}{\alpha}

\newcommand{\vak}{\varkappa}
\newcommand{\pa}{{\partial}}

\newcommand{\si}{\sigma}
\newcommand{\Si}{\Sigma}

\newcommand{\fr}{\frac}

\newcommand{\bw}{\begin{widetext}}
\newcommand{\ew}{\end{widetext}}
\newcommand{\be}{\begin{align}}
\newcommand{\ee}{\end{align}}
\newcommand{\ba}{\begin{eqnarray}}
\newcommand{\ea}{\end{eqnarray}}

\newcommand{\ep}{\epsilon}

\def\cV{{\cal V}}

\def\e{{\rm e}}

\newcommand{\nul}{\, ^0  }
\newcommand{\un}{\, ^1  }
\newcommand{\de}{{\, ^2}  }
\newcommand{\pp}{\, ... \,}

\newcommand{\ret}{\rm ret}

\newcommand{\loc}{\rm loc}
\newcommand{\nloc}{\rm nloc}
\newcommand{\reg}{\rm reg}

\newcommand{\sgn}{{\rm sgn\,}}
\newcommand{\nn}{\nonumber}

\newcommand{\vp}{\vphantom{\frac{a}{a}}}

\newcommand{\kn}{{\kern 1pt}}
\newcommand{\akn}{{\kern -1pt}}

\newcommand{\ah}{ \mathrm{a}}
\newcommand{\bh}{ \mathrm{b}}

 \numberwithin{equation}{section}

\begin{document}

\title{ Branestrahlung: Radiation in the particle-brane collision}
%%%%%  Authors  %%%%
\author{Dmitry Gal'tsov, Elena Melkumova and Pavel
Spirin
\thanks{\tt E-mail: galtsov@physics.msu.ru, elenamelk@mail.ru
pspirin@physics.uoc.gr}}
\affiliation{Faculty
of
Physics, Moscow State University, 119899, Moscow, Russian Federation.}

\pacs{11.27.+d, 98.80.Cq, 98.80.-k, 95.30.Sf}
\date{\today}

\begin{abstract}
We calculate the radiation accompanying  the gravitational collision of
a domain wall a the point particle in five-dimensional
spacetime. This process -- which can be regarded as brane-particle
bremsstrahlung, here called {\it branestrahlung} -- has  unusual
features. Since the brane has intrinsic dynamics, it gets excited in
the course of collision, and, in particular, at the moment of
perforation the shock branon wave is generated, which then expands
with the velocity of light. Therefore, apart from the time-like
source whose radiation can be computed in a standard way, the total
radiation source contains a light-like part whose retarded field is
quite nontrivial, exhibiting interesting retardation and memory
effects. We analyze this field in detail, showing that -- contrary to
the claims that the light-like sources should not radiate at all --
the radiation is nonzero and has classically divergent spectrum. We
estimate the total radiation power introducing appropriate cutoffs.
In passing, we explain how the sum of the nonlocal (with the
support inside the light cone) and the local (supported on the cone)
singular parts of the Green's function of the five-dimensional
d'Alembert equation together defines a regular functional.
\end{abstract}
\maketitle

%\tableofcontents

\setcounter{equation}{0}
\section{Introduction}
The gravitational interaction of domain walls with matter was
extensively discussed in the past in the context of four-dimensional
cosmology \cite{linear,linear1,Vilsh}.  More recently, the general
development of the theory of branes in the superstring setting
stimulated investigation of other aspects of this interaction, such
as collisions  of domain walls and other branes with black holes
\cite{Flachi:2007ev,BBHS4,BBHS5}. Especially interesting is the
process of perforation, which was suggested as a new mechanism of
domain-wall destruction in the early Universe
\cite{ChamEard,Stojkovic:2005zh,Flachi:2007ev}. Other applications
refer to the braneworld models of the Randall-Sundrum type
\cite{RS1,RS2,Tanahashi:2011xx,BBHS3}, in particular,  the problem
of escape of black holes from branes \cite{escape,escape1,escape2}.

In most of the literature the brane-black hole problem is
treated by considering one object as creating the fixed
background, and another as moving in this background.
But genuine two-body features of their interaction including
the backreaction are very difficult to describe in the full
nonlinear setting. In an attempt to explore such aspects of
the interaction, we consider a simplified model in which the
second body is the point-like mass living in the bulk.
Recently we have investigated the perforation of a domain
wall by a point mass in a particularly simple setting of
the linearized gravity~\cite{GaMeS1,GaMeS2}.
 Due to the continuous nature
of the gravitational field of the domain wall in its location,
one can describe the perforation of the brane in terms of
distributions. This approach allows one to consider one
novel aspect that is difficult to analyze in the full nonlinear
theory, namely, the excitation of the internal degrees of
freedom of the domain wall in the course of collision. The
domain wall gets excited after the perforation in the form of
the spherical shock branon wave propagating outwards
along the brane with the speed of light. This wave is a
reaction of the wall to the change of the particle's acceleration
 by a finite amount at the moment of piercing. Thus,
contrary to the two-particle case, the collision becomes
classically nonlocal. The energy-momentum balance in this
process can therefore be defined only instantaneously, as it
takes into account the contribution of the branon wave \cite{GaMeS2}.

Here we calculate the radiation accompanying the
particle-–domain-wall collision which is somewhat similar
to bremsstrahlung. Actually we are interested in gravitational
radiation, but in order to simplify the computation we
consider the emission of the scalar waves, introducing the
interaction of the massless scalar field with the trace of the
brane energy-momentum tensor. We realize that the spectral
distribution of radiation is essentially dependent on its
spin, so the scalar case cannot fully reproduce gravitational
radiation. However, such aspects as the propagation of the
wave fronts and causality structure of the radiation field
should be similar for all spins, and in this paper we are
interested mainly with this. The novel feature of this
radiation problem is the presence of the light-like source
due to the branon as a part of the total source.
Recently, the retarded solutions of the d'Alembert equation with light-
like sources were extensively discussed in connection with
the radiation problem \cite{Azzurli:2012tp,Azzurli:2014lha},
the memory effect
\cite{Tolish:2014bka,Tolish:2014oda,Bieri:2013ada,Winicour:2014ska,Bieri:2015yia}
and the Bondi-Metzner-Sachs asymptotic symmetries
\cite{Strominger:2014pwa}. Our radiation problem provides a novel interesting setting
for these studies.

 As far as
the problem of radiation from massless sources is concerned, the
arguments were presented that radiation in this case is totally absent
 \cite{Lechner:2014kua,Kosyakov:2007pm}. Indeed, within the
classical theory the retarded potentials from massless point sources
diverge on a line parallel to the velocity, and various ways to
regularize these divergences —- which also have quantum
counterparts as collinear
 divergences~\cite{Kinoshita:1962ur,Lee:1964is,Nauenberg:1965uka,Weinberg:1965nx} --
were suggested. An ultimate result of such procedures  was sometimes
claimed to be zero. Meanwhile the relativistic Li\'{e}nard formula
for the total radiation power from an accelerated point charge diverges
in the massless limit. This controversy was resolved in \cite{SR},
where it was shown that radiation
from the massless charge is nonzero and finite, but (being
essentially quantum) it is described by classical theory only
in the IR part of the spectra.
 We show that
{\it branestrahlung} -- which always contains a contribution from the massless
component of the source—-also has a similar property
requiring appropriate cutoffs.

The computation of radiation depends on the spacetime
dimension in a nontrivial way, being different, in particular,
in the case of even and odd dimensions~\cite{GS}. Here we
restrict ourselves to the case of the five-dimensional bulk,
appealing primarily to the Randall-Sundrum II (RS\,II) kind
of setting. Radiation in five dimensions does not satisfy
Huygens' principle which leads to additional new features
as compared with the four-dimensional case. In particular,
we discuss the structure of the retarded Green's function of
the five-dimensional d'Alembert operator whose treatment
is often confusing in the literature, emphasizing the
importance of its purely local part as a regulator.

\setcounter{equation}{0}
\section{The Setup}
We consider a  $3$-brane embedded in the $5$-dimensional  spacetime
(the bulk) with the  metric $g_{MN}$, where $M,N=0,1,2,3,4$.  The
brane worldvolume $\cV_{3+1}$  defined by the embedding equations $
x^{M}=X^{M}(\sigma_\mu)\; $ is parametrized by the coordinates
$\sigma_\mu,  \;
    (\mu=0,...,3)$ on $\cV_{3+1}$ .
The brane gravitationally interacts with a point mass $m$  moving
in the bulk.  In addition, the brane(and not the particle)is coupled with the bulk scalar field
 $\varphi$, being endowed with the charge density $f$. Therefore the particle-brane interaction is purely
 gravitational, while the scalar radiation is generated only by
 the brane. This simplifies the problem considerably, while
still keeping the main features of the realistic gravitational radiation.

The corresponding action  reads
\begin{align}\label{Bac}
S = -\fr{\mu}{2}\int\left[\vp\,
  X_\mu^M X_\nu^N g_{MN}\gamma^{\mu\nu}-2\right]\left(1+\frac{f}{\mu}\,\varphi\right)\sqrt{-\gamma}
  \,d^{\kn 4}
  \sigma  - \frac{1}{2} \int \! \left(e\kn\kn    g_{MN}\dot{z}^M
\dot{z}^N+\frac{m^2}{e}\right) d\tau -\nn \\
-\frac{1}{\vak^2_5}\!\int\! R  \,\sqrt{g}\; d^{\kn 5} x -\frac{1}{2}
\int\! g_{MN} \nabla^{M} \varphi\, \nabla^{N}\varphi \,\sqrt{g}\;
d^{\kn 5} x \,.
 \end{align}
Here $\mu$ is the brane mass density (tension), $f$ is the brane
scalar density, $X_\mu^M=\pa X^M/\pa\hsp\sigma^\mu$ are the brane
tangent vectors,   $g={\rm det}\, g_{MN}$ and $\gamma^{\mu\nu}$ is
the inverse metric on $\cV_{3+1}$, with $\gamma={\rm det}
\,\gamma_{\mu\nu}$. Also $\vak^2\equiv 16\pi G_5$, $e(\tau)$ is the
einbein on the particle worldline, the metric signature is
$(+----)$, and we use the Landau-Lifshitz convention for the curvature.
Note that due to the choice of the signature, in $D=5$ one has $g>0
$, while $\gamma<0$.

It is worth noting that by taking the interaction of the
scalar field with gravity into account we introduce a
nonlinearity similar to that described by the three-graviton
vertex in the genuine gravitational problem. Due to this
interaction, the effective radiation source term will include
the stress part in the same way that gravitational stresses
contribute to the gravitational radiation amplitude as
prescribed by the Bianchi identity.

The first term in Eq.(\ref{Bac}), representing the action of the brane,
has to be varied independently over $X^M$ and $\gamma^{\mu\nu}$   to
get the equations of motion and the constraint. Similarly, the
particle  term in Eq.(\ref{Bac})  has to be varied over $z^M$ and $e$,
while the terms containing the bulk fields $\varphi$ and $g_{MN}$,
before being varied over them, have to be extended to the bulk
integrals by inserting the appropriate delta functions.

The variation of Eq.(\ref{Bac}) with respect to $X^M$ and $\gamma^{\mu\nu}$
gives the brane equation of motion in the covariant form
  \begin{align} \label{em_red}
\nabla_\mu\akn\akn \left( \vp X_{\nu}^N
g_{MN}\gamma^{\mu\nu}\sqrt{-\gamma}\,(\mu+ f
\varphi)\right)=\sqrt{-\gamma}\smhsp f \varphi_{,M}\,.
 \end{align}
 with $\nabla_\mu \equiv X^M_\mu \nabla_M$, and the constraint equation
  \begin{align}\label{im}\left.\gamma_{\mu\nu}=X_\mu^M
X_\nu^N g_{MN}
 \right|_{x=X}\,,\end{align}
where we define $\gamma_{\mu\nu}$ as the induced metric on
$\cV_{4}$.
 Varying $S$ with respect to $z^M(\tau)$ and $e(\tau)$
one obtains the corresponding system for the particle:
\begin{align}\label{eomp}
&\fr{d}{d\tau} \left( e\, \dot{z}^N g_{MN}  \right)=\frac{e}{2} \;
g_{NP,M}\,\dot{z}^N \dot{z}^P ,\\
&\label{consp} e^2  g_{MN} \dot{z}^M \dot{z}^N=m^2\,.
\end{align}
Finally, the variation over $\varphi$ leads to the scalar field
equation
\begin{align}\label{Eeq}
g_{MN} \nabla^M \nabla^N \,\varphi=\rho\,,
\end{align}
with the source
\begin{align}\label{rho_generic}
\rho=\frac{q}{2}\,\int \left[\vp\,
  X_\mu^M X_\nu^N g_{MN}\gamma^{\mu\nu}-2\right]
\sqrt{-\gamma}\; \hsp \frac{\delta^{5}\!\!
\left(x^{M}-X^{M}(\sigma^{\nu})\right)}{\sqrt{g(x)}}\;d^{\kn
4}\sigma\,,
\end{align}
or, substituting  Eq.(\ref{im}),
\begin{align}\label{Eeq1}
\rho\smhsp(x)=f\int \sqrt{-\gamma}\;\hsp
\frac{\delta^{5}\!\left(x^{M}-X^{M}(\sigma^{\nu})\right)}{\sqrt{g(x)}}
\; d^{\kn 4}\sigma \,.
\end{align}
\subsection{Iteration scheme}
Next we present the bulk metric as the perturbed Minkowski metric
$\eta_{MN}$
\begin{align}
\label{meka} g_{MN}=\eta_{MN}+ \vak \hsp H_{MN}\,,
\end{align}
and expand  all quantities in powers of $H_{MN}$, using $\eta_{MN}$
to raise and lower the indices. As usual, we   impose  the
flat-space harmonic gauge
\begin{align}
\label{hagef} \pa_N H^{MN}=\frac{1}{2} \pa^M H \,, \qquad H \equiv
\eta^{LR} H_{LR}\,.
\end{align}

Our technique consists in solving  the equations for the bulk
metric, the embedding functions $z^M(\tau),\; X^M(\sigma^\mu)$, the
Lagrange multiplier $e$, the worldvolume metric $\gamma_{\mu\nu}$,
and the scalar $\varphi$ by iteratively expanding them in terms of the
couplings $\vak, \;f$. When doing this we have to keep only the mutual
interaction terms and omit the self-action. The goal is to compute the
scalar radiation in the lowest reliable order in both couplings.

 The zero-order solution is trivial. It describes the free flat
unperturbed brane and the particle moving with constant velocity
normally to the brane $u^M=\gamma(1, 0,0,0, v)$, where
$\gamma=1/\sqrt{1-v^2}$. Correspondingly, the trajectory $\nul
z^M=(\nul t (\tau), 0,0,0, \nul z(\tau))$  is
   the straight line
$ \nul z^M(\tau)= u^M\tau $. The einbein  is chosen equal to the
particle mass $\nul e =m $, so that the trajectory is parametrized
by the  proper time and the velocity satisfies the normalization
$\eta_{MN}u^M u^N= 1$.

In the zeroth order in $\vak$ the brane is assumed to be unexcited
and to fill the plane $z=0$,
 \begin{align}
\nul \!X^M(\si) =\Si^M_\mu \si^\mu\,,\qquad
\Si^M_\mu\Si^N_\nu\eta_{MN}=\eta_{\mu\nu}\,,
 \end{align}
so the internal coordinates on the brane coincide with the
corresponding bulk coordinates and the induced metric on $\cV_{4}$
is flat: $\gamma_{\mu\nu}=\eta_{\mu\nu}$.  Obviously, this is a
solution to  Eq.\,(\ref{im}) for $\vak=0$, and  it is convenient
to fix $\Si^M_\mu=\delta^M_\mu$ without loss of generality. In other
terms, we choose the Lorentz frame where the unperturbed brane is at
rest.

In the first order the metric deviation is the sum of the contribution
of the brane and the particle:
\begin{align}
\un  H_{MN}= h_{MN} +  \bar{h}_{MN}\,,
\end{align}
which satisfy
 \begin{align}\label{bre0ss}
  \Box\, h^{MN}=- \vak \left(\nul
T^{MN}-\frac{1}{3} \nul T \eta^{MN}\right), \qquad
 \Box\, \bar{h}^{MN}=- \vak \left(\nul \bar{T}^{MN}-\frac{1}{3}  \nul
\bar{T}\, \eta^{MN}  \right)\,,
\end{align}
with  $\Box \equiv \pa_M\pa^M$ being the flat five-dimensional
d'Alembert operator and $\nul T \equiv \nul T^{MN} \eta_{MN}$. The
source terms read
 \begin{align}\label{bre0}
&\nul T^{MN}=\mu\int \Si^M_\mu\Si^N_\nu
\eta^{\mu\nu}\,\delta^{\kn 5}\!\left(x-\nul{\! X} (\si)\right)\,
d^{\kn 4}\si \,,\\ &\label{T0mn} \nul \bar{T}^{MN}=  m\int u^M u^N
\delta^{\kn 5}\!\left(x-\nul z(\tau)\right)\, d\tau\,.
 \end{align}

Using $h^{MN}$ and the zeroth order quantities in Eqs.\,(\ref{consp}) and (\ref{eomp}) one obtains for $\un e$ and $\un z^M$
the equations\,\footnote{Our gauge condition is $g_{MN} \dot z^M \dot
z^N=1$. To this order it reduces to $^1e=0$.}
\begin{align}
\label{e1eq} \un e=-\frac{ m}{2} \left( \vak
h_{MN}u^Mu^N+2\,\eta_{MN}u^M \un
 \dot{z}^N \vp \right)
\end{align}
and Eq.(\ref{eomp}) (upon elimination of $\un e$) give for $\un
z^M$
\begin{align}\label{z1eq}
 \fr{d}{d\tau} \un\dot{z}_M=-\vak
 \left(h_{PM,Q}-\fr12\, h_{PQ,M}  \right) u^P u^Q \,.
\end{align}

Substituting the first-order metric deviation $\bar{h}_{MN}$ due to the
particle (to be determined below) and the first-order brane
perturbations into Eq.(\ref{Eeq1}),  and computing the perturbation of the
induced metric (\ref{im}), $\gamma_{\mu\nu}=\eta_{\mu\nu}+\un
\gamma_{\mu\nu}+ \mathcal{O}\hsp(\vak^2 h^2)$,
 \begin{align}\label{indmetrvar}
 \un \gamma_{\mu\nu}=2\hsp \delta^M_{(\mu}\un{\!X}^N_{\nu)}\eta_{MN}+
 \vak \smhsp \bar{h}_{\mu\nu}\,,\qquad  \un \sqrt{-\gamma} =
 \frac{1}{2}\left( 2\hsp \delta^M_{\mu}\un{\!X}^N_{\nu}\eta_{MN} \eta^{\mu\nu} +
 \vak \smhsp \bar{h}^{\lambda}_{\lambda}\vp\right)\, ,
 \end{align}
 one gets
\begin{align}  \label{pisk}
 {\Pi}_{MN}\;\Box_{D-1}\;\un{\!X}^N= {\Pi}_{MN}\;J^N\,, \qquad  {\Pi}^{MN} \equiv \eta^{MN}-
\Sigma^{M}_{\mu}\Sigma^{N}_{\nu} \eta^{\mu\nu}\,,
 \end{align}
where $\Box_{D-1} \equiv \partial_{\mu} \partial^{\mu}$ and $
{\Pi}^{MN} $ is the projector onto the (one-dimensional) subspace
orthogonal to $\cV_{D-1}$. The source term   in  Eq.(\ref{pisk}) reads
 \begin{align} \label{JN}
 J^N=   \vak \, \Sigma_P^\mu \,\Sigma_Q^\nu\,
\eta_{\mu\nu} \left(\frac{1}{2} \, \bar h^{PQ,N} -\bar
h^{NP,Q}\right)_{\!z=0}\!,
 \end{align}
where ${\rule{0em}{0.8em}\Sigma}^{\al}_{\nhsp M} \equiv
\Sigma^{N}_{\nu} \eta^{\nu \al} \eta_{MN} $.

Finally, the equation for $\un \varphi$ is given by the leading
order of Eq.(\ref{Eeq}):
 \begin{align}\label{unphieqn}
  \Box\, \un \varphi = \nul \rho=f\int
 \delta^{4}\!\left(x^{\mu}-\sigma^{\mu}\right)\,\delta(z)
\; d^{\kn 4}
  \sigma =f\,\delta(z)\,,
 \end{align}
with $\nul \rho$ being the leading order of Eq.(\ref{Eeq1}).

Equations (\ref{bre0ss}), (\ref{e1eq}), and (\ref{z1eq}), together
with an equation for the brane perturbations, form a complete set of
equations to this order. The gauge fixing condition (\ref{hagef})
for $\un \psi^{MN}$ is a consequence of the conservation of $\nul
T^{MN}$. To solve the d'Alembert equation for the field variables we
will use the Fourier transforms  defined as
\begin{align}\label{sc_pert3}
\varphi(x)=\frac{1}{(2 \pi)^5}\int \varphi(k) \, \e^{-i kx} d^{\kn 5}\nhsp k \,,\qquad \varphi(k)= \int \varphi (x) \,\e^{i kx} d^{\kn 5} \nhsp x\, ,
\end{align}
where $kx\equiv k_M x^M$.
 \subsection{Radiation}
We will compute the scalar radiation generated in the
 particle-brane collision as the simplified version of the
gravitational radiation. Recall that the particle is assumed to
not carry the scalar charge, so the interaction between the
brane and the particle is purely gravitational. Gravity also
interacts with the scalar field, which is generated by the
brane and lives in the bulk. The radiative component of the
scalar field, which is detected by the nonzero Fourier
transform
$\varphi(k)$ on the mass shell $k^2=0$ (five-dimensional
square), appears at the lowest order in the second iteration,
namely, the first in $\vak$ and the first in $f$. Expanding Eq.~(\ref{Eeq}) in
$\vak$, we obtain in this order
\begin{align}
\label{psi2eq} \Box\, \de\varphi=\vak\;j\,,
\end{align}
with the source term consisting of two contributions,
\begin{align}\label{source2}
j=   \tilde{\rho}+S\,, \qquad\qquad S \equiv
-\partial_M\!\left(\frac{1}{2}\,h\,
\partial^{M}-h^{MN}
\partial_N\right)\!\! \un\varphi\, ,
\end{align}
while $\tilde{\rho} \equiv \un ( \rho \sqrt{g})$ is the direct, or
{\textit{local},  current,
\begin{align} \label{taumn}
   \tilde{\rho}  =
\fr{f}{2} \int & \left[\,  \vak \smhsp
\left(\bar{h}^{\lambda}_{\lambda} -\bar{h}^{L}_{L}
\right)+2\un{\!X}^{L}_{\lambda}\Sigma^{\lambda}_L -2\un{\!X}^{L}
\partial_L \vp \right]\delta^5\!\left(x^A- \Sigma^A_\al\sigma^\al\right)\, d^{\kn 4}\si \,,
 \end{align}
 whereas the nonlocal term $S$,  (which depends on the bulk
gravitational field) is due to the expansion of the curved-
space d'Alembert operator in $\vak$.

The total momentum loss (for a more detailed analysis of the momentum
balance in our problem see Ref.~\cite{GaMeS2}) in the collision is
computed in the standard way and is expressed in terms of the
Fourier transform of the current as follows:
\begin{align}\label{sc_pert9}
\Delta P^{M}=  \frac{1}{(2 \pi)^{4}}
 \int \theta(k^0)\,
 k^{M} \, \delta(k^2)\, |j (k)|^2 \, d^{\kn 5} \akn k\, .
\end{align}
Finally, introducing the frequency $\omega=k^0$ and integrating over
$|\mathbf{k}|$,  for $E_{\rm rad}=\Delta P^{0}$ we arrive at
\begin{align}\label{sc_pert10}
E_{\rm rad}= \frac{1}{2\smhsp(2 \pi)^{4}}
\int\limits_{0}^{\infty}\omega^{3} d\omega \int\limits_{S^{3}}
d\Omega \;
 |j (k)|^2\,,
\end{align}
where  the wave five-vector $k$ is  null. We will parametrize it by
the frequency, the normal to the brane projection $k^z$ and the
 three-dimensional  vector ${\bf k}_\bot$, parallel to the brane:
\begin{align}
k^2=k^M k_M=\omega^2 -(k^z)^2 - (k_\bot)^2=0\,.
\end{align}

\setcounter{equation}{0}
\section{The first order}
The first-order solutions are obtained  straightforwardly by  passing
to the momentum space, so we just list the corresponding results.
\subsection{Linearized fields}
The brane gravitational field $h_{MN}(q)$  is obtained by
taking the Fourier transform of the energy-momentum
tensor (\ref{bre0}) and
dividing by the box operator:
 \begin{align} \label{ge2}
  h_{MN}(q)=  \frac{ (2\pi)^{4}\vak \smhsp \mu  \,\delta^{4}(q^{\mu}) }{q^{2}}
   \left( \delta_M^\mu \delta_{N}^{\nu} \eta_{\mu\nu}-\frac{4}{3}\,\eta_{MN}\right).
 \end{align}
In the coordinate representation it reads
 \begin{align} \label{brgr}
h_{MN}(x)=\frac{\vak \mu}2\left(\hsp \delta_M^\mu \delta_{N}^{\nu}
\eta_{\mu\nu} -\frac{4}{3}\,\eta_{MN}\right)\!|z|=\frac{\vak \mu
|z|}{6}\:{\rm diag}\hsp(-1,1,1,1,4)\,,
 \end{align}
with the trace being
$$ h=-\frac{4}{3} \,\vak \mu |z|\,.$$

The solution of the linearized  equation (\ref{unphieqn}) for the
scalar field in the momentum representation is
 \begin{align} \label{ge2s}
 \un \varphi(q)= - \frac{ (2\pi)^{4}  \smhsp f \,\delta^{4}(q^{\mu}) }{q^{2}}\,.
 \end{align}
The corresponding coordinate-space solution
 \begin{align} \label{ge2A}
 \un \varphi(x)=  - \frac{1}{2} f |z|
 \end{align}
is proportional to the trace of the gravitational perturbation.

Similarly, for the particle's metric perturbation we get
 \begin{align}\label{ge_mom}
   \bar{h}_{MN}(q) =\frac{2\pi \vak\smhsp m \smhsp \delta(qu)}{q^2+i \varepsilon q^0} \left(u_M
u_N-\fr{1}{3}\,\eta_{MN}\right)\,,
  \end{align}
or, in the coordinate representation,
\begin{align} \label{hpart}
 \bar{h}_{MN}(x)=-\frac{\vak
\smhsp m }{4\pi^{2}}
 \left(u_M
u_N-\frac{1}{3}\,\eta_{MN}\right)\frac{1}{ \gamma^2(z-v t)^2+r^2}\,,
 \end{align}
where $r^2= \sum_{i=1}^n(\sigma^i)^2$. This is nothing but the
Lorentz-contracted $D$-dimensional Newton gravitational field of a
uniformly moving point particle.
\subsection{Branon}\label{Branon_wave_equation}
The picture of the collision looks as follows. The particle impinges
on the brane normally and perforates it at the moment $t=0$.
The particle's gravitational field causes the
perturbation of the brane worldvolume, which is somewhat
nontrivial to analyze (the details of which can be found in
Refs.~\cite{GaMeS1,GaMeS2}\kn\footnote{The transverse
coordinate of the brane can be viewed as the Nambu-Goldstone boson
(branon) which appears as a result of spontaneous breaking of the
translational symmetry \cite{KuYo}. In the braneworld setting it is
coupled to bulk gravity and matter on the brane via the induced
metric~\cite{Bu}.}); here, we just present the results. Only the
perturbation of $X^N$ transverse to the brane is physical; the
longitudinal ones can be gauged away by the coordinate
transformation in the worldvolume. Denoting this component as
another scalar (branon) $\Phi(x^\mu)=\un {\!X}^z$ and linearizing
the Nambu-Goto equation, we obtain for the branon the
four-dimensional d'Alembert equation
 \begin{align} \label{NGEQ}
\Box_{4} \Phi(\si^{\mu})=J(\si^{\mu}),
  \end{align}
with the source term
\begin{align} \label{jxb}
J( \sigma )=\vak \left[\frac{1}{2}\,  \eta_{{\mu\nu}}\bar{h}^{\hsp
\mu\nu,z}-\bar{h}^{\hsp  z\hsp  0,0}\right]_{z=0} =-  \fr{\la
vt}{[\gamma^2 v^2 t^2+r^2]^2}\,,\qquad \la=\fr{\vak^2 m\gamma^2
}{4\pi^{2}}\left( \gamma^2 v^2 +\fr{1}{3}\right) .
 \end{align}
In the
  momentum space the retarded solution to this d'Alembert equation
  reads
 \begin{align} \label{NGEQ_mom}
 \Phi(q^{\mu})=- \frac{ J(q^{\mu})}{q_{\nu}q^{\nu}+2i\ep q^0}\,,
 \qquad J(q^{\mu})=-\frac{2\pi^{2}\la\smhsp }{\gamma
 }\frac{i q^0}{\gamma^2 v^2
\mathbf{q}^2+(q^0)^2}\,.
  \end{align}
Expanding the double pole into the sum of two simple poles, we get
the decomposition  $\Phi \equiv \Phi_\ah+\Phi_\bh$, where
 \begin{align}
 \Phi_\ah(q^{\mu})=-\frac{2\pi^{2}\la i}{q^0}
 \frac{\gamma^2 v^2}{\gamma^2 v^2
\mathbf{q}^2+(q^0)^2}\,,\qquad \Phi_\bh(q^{\mu})=\frac{2\pi^{2}\la
i}{q^0}
 \frac{1}{q^2+2i\ep q^0}\,,
  \end{align}
which have different physical meanings. The component $\Phi_\ah$,
which in the coordinate representation reads
\begin{align} \label{Phia}
\Phi_\ah=- \frac{
  \la}{2  \gamma^3 r}  \,\arctan \frac{r}{\gamma v  t }\, ,
\end{align}
 is \textit{antisymmetric} in time. Recall that the gravitational
force between the brane and a particle is repulsive and it changes
sign at the moment of piercing. The brane gets deformed in
the direction opposite to the instantaneous location of
particle, and this deformation (described by $\Phi_\ah$) is rigidly
tight to the particle motion, without retardation. In the
momentum space it is due to the pole on the imaginary axis.

The second component $\Phi_\bh$ is the shock branon wave which is excited just at the moment of perforation. It arises
from the pole on the real axis and corresponds to a (quasi)
free branon wave propagating with the velocity of light
outwards from the center
$r=0$ where the particle perforates the brane:
\begin{align} \label{Phibg}
\Phi_{\bh}=   \frac{
 \pi  \la}{ 2 \kn r \gamma^3} \, \theta(t)\,\theta(r-t)\,.
\end{align}
What is more surprising is that it is only quasifree, since by
acting with the d'Alembert operator one finds the derivative
of the delta function:
 \begin{align}\label{jajcc2}
\Box_{4}{\Phi}_{\bh}  = \frac{
 \la  \, \pi }{ 2\gamma^3
r }\, \delta'(t)\,.
\end{align}
The peculiar nature of this source is explained as follows.
The force between the particle and the brane at the moment
of perforation is nonzero, and it changes sign. It acts as the
singular local flush which causes the excitation of the shock
branon wave. It is important that the shock wave now
becomes a source of radiation, i.e., we have to explore the
retarded solution of the bulk d'Alembert equation with such
a light-like source. This is a rather nontrivial problem,which
we solve in the next section and the Appendix. It leads to a
complicated structure of fields propagating in the bulk.

The ``rigid'' part of the deformation of the brane (\ref{Phia})
satisfies the following equation:
 \begin{align}\label{jajcc3}
\Box_{4}{ \Phi}_{\ah} = -\frac{
 \la  \,\pi }{ 2\gamma^3
r}\, \delta'(t)- \frac{\la vt}{[\gamma^2 v^2 t^2+r^2]^2}\,,
\end{align}
so in the sum of both contributions the delta-function sources
cancel as expected.

In the ultrarelativistic limit the shock wave dominates. The second
term at the right-hand side of the Eq.~(\ref{jajcc3})  in this
case behaves like the regularized derivative of the delta function
as $\gamma\to\infty$; an  accurate calculation shows that it exactly
cancels the first term. So, the direct action of the particle upon
the brane vanishes for a light-like particle. This is not surprising
since the infinite boost causes the gravitational field of the
particle to become a shock wave acting solely at the moment of
perforation. Hence the total branon solution caused by the
light-like particle is determined by $\Phi_{\bh}$ only. In this case
one has to use as a parameter the particle energy $\mathcal{E}$,
arriving at
$$ \Phi=   \frac{\vak^2 \mathcal{E}
}{8\pi }   \, \frac{\theta(t)\,\theta(r-t)}{r} \,.$$

\setcounter{equation}{0}
\section{Retarded field of the light-like branon source}\label{phi2}
Before proceeding with the calculation of the total radiation amplitude
$\de\varphi$ generated by the current $j(k)$, we consider its
relativistic component $\psi$ generated by the shock branon part of
the source term. As we have seen, this part becomes an exact field
in the case of a particle of zero mass. The construction of the
retarded solution of the d'Alembert equation for the light-like
source is nontrivial, especially in odd spacetime dimensions since
the Green's function is not localized on the light cone.

Thus we seek the retarded solution of the following five-dimensional
wave equation:
\begin{align} \label{Phib}
\Box\psi= - \alpha\, \frac{\theta(t)\,\theta(r-t)}{r}\, \delta'(z)\,,
\end{align}
where we defined $\alpha= { \pi \la f}/{ 2 \gamma^3}$. We will
abbreviate the product of two Heaviside functions as
$$\Theta_x(a,b) = \theta(x-a)\,\theta(b-x)  =
\left\{%
\begin{array}{ll}
    0, & \hbox{$x<a$;} \\
    1, & \hbox{$a\leqslant x<b$;} \\
    0, & \hbox{$x\geqslant b$.} \\
\end{array}%
\right.\qquad\qquad \text{for } b>a\,. $$ In this  notation
$\theta(t)\,\theta(r-t)=\Theta_{t}(0,r)$.  This double Heaviside
function restricts the domain of integration to a finite interval: $
\int \Theta_x(a,b) f(x, \pp)\,dx = \int_a^b f(x, \pp)\,dx \,.$

\subsection{Regularization of the Green function in five dimensions}
The retarded Green's function in five dimensions satisfying the
equation $\Box G=-\delta^5(x-x')$ is obtained by differentiation of
the three-dimensional Green's function localized inside the light
cone~\cite{GS} (recall again that Huygens' principle does not
hold in odd spacetime dimensions):
\begin{align} \label{Gre0}
G^{\ret}(X)=-\frac{ \theta(T)}{2\pi^2}\,  \frac{d}{dX^2}
\frac{\theta(X^2)}{(X^2)^{1/2}}, \qquad\qquad X^M=x^M-{x'}^M.
\end{align}
One obtains therefore the sum of the nonlocal ($\sim\theta$) and the
local ($\sim\delta$) terms:
\begin{align} \label{Gre1}
\ds G^{\ret}(X)=\frac{1}{4\pi^2}
\left(\frac{\theta(T-R)}{(T^2-R^2)^{3/2}}-
\frac{\delta(T-R)}{R(T^2-R^2)^{1/2}}
 \right)\,,\qquad R^2 \equiv
 (\mathbf{r}-\mathbf{r}')^2+(z-z')^2\,.
\end{align}
As $\chi\equiv T\pm R\to +0$ these terms contain nonintegrable
singularities $\chi^{-3/2}$ and $\delta(\chi)/\chi^{1/2}$
respectively. Nonetheless, their sum represents the \textit{regular}
functional. To see this we apply the regularization  $T\to T-\epsilon$
assuming that the limit $\epsilon \to +0$ has to be taken after
summing up both contributions.  The regularization shifts the
support of the Green's function from the interior of the future
light cone $T^2-R^2=0$ into the half of the time-like hyperboloid
$T^2-R^2=\epsilon^2$ with an apex at $x^M$. The distance between the
cone and the hyperboloid measured along the $T$ axis is $\epsilon$.
With this regularization,
\begin{align} \label{Gre2}
\ds G^{\ret}_{\epsilon}(X)=\frac{1}{4\pi^2}\,  \lim_{\epsilon \to
+0} \left(\frac{\theta(T-R-\epsilon)}{(T^2-R^2)^{3/2}}-
\frac{\delta(T-R-\epsilon)}{R(T^2-R^2)^{1/2}}
 \right)\,,
\end{align}
where $T>0$ and $R>0$ are assumed. We are going to show that the
divergent parts of the  two terms mutually cancel and after removing
the regularization we are left with the finite result.

Thereby we have to compute
 \begin{align} \label{ta5}
 \psi=\frac{1}{4\pi^2}\, \lim_{\epsilon \to +0} \int
\left(\frac{\theta(T-R-\epsilon)}{(T^2-R^2)^{3/2}}-
\frac{\delta(T-R-\epsilon)}{R(T^2-R^2)^{1/2}}
   \right) \frac{\theta(t')\,\theta(r'-t')}{r'}  \:\delta'(z') \,d^{\kn 5} x'\,.
 \end{align}
Integrating over $z'$ with the derivative of the delta function, one
can pass a differentiation over $z$ since the integrand depends on
$z'$ and $z$  through the difference $z-z'$, obtaining
\begin{align} \label{ta6d}
 \psi= - \alpha \,\frac{\partial J}{\partial z}\,,
 \end{align}
where
\begin{align} \label{ta6}
J \equiv   \lim_{\epsilon \to +0} \int
\left(\frac{\theta(T-R-\epsilon)}{(T^2-R^2)^{3/2}}-
\frac{\delta(T-R-\epsilon)}{R(T^2-R^2)^{1/2}}
   \right) \frac{\Theta_{t'}(0,r')}{r'}  \: \,d^{\kn 4} x'\,
 \end{align}
 and, from now  on, $R \equiv \sqrt{
(\mathbf{r}-\mathbf{r}')^2+z^2}$. Since the source is nonzero only
for $t'>0$ and $R\geqslant 0$, it follows that the solution is
nonzero at $t>0$ as expected.

First we consider  contribution of the local part of the Green's
function (\ref{ta6})by  integrating over $t'$ with the delta function.
If  $t-R$ lies outside the interval $(0,r')$, the integral vanishes;
otherwise,
 \begin{align} \label{ta9}
J_{\hsp\loc} = -  \int \frac{\theta(t-R)\,\theta(r'-t+R)  }{
R(2R+\epsilon)^{1/2}  r' \sqrt{\epsilon}}    \;d^{\kn 3} \mathbf{r}'\,,
 \end{align}
which diverges as   $\epsilon \to +0$.

The  nonlocal term in Eq.~(\ref{ta6}) is more complicated. Using the
identity
\begin{align} \label{ta13}
\theta(r'-t')\,\theta(t-t'-R-\epsilon) =
    \theta(r'-t')\, \theta(t-R-r')+
    \theta(t-t'-R-\epsilon)\,\theta(r'-t+R)\,,
 \end{align}
we split it into two parts obtaining after integration over $t'$
\begin{align} \label{ta16}
J_{\nloc}=\int
   \left[  \left(\vp F(r')-F(0)\right)\, \theta(t-R-r')+
    \left(\vp F(t-R-\epsilon)-F(0) \right)\,\theta(r'-t+R)\,\theta (t-R)\right]\,  \frac{d^{\kn 3} \mathbf{r}'}{r'},
 \end{align}
where the function
\begin{align} \label{ta17}
F(x)=\frac {t-x}{ R^2\,\sqrt {(t-x)^{2} -R^2}}
 \end{align}
is an antiderivative of $1/[(t-x)^2 -R^2]^{3/2}$. Consider now the
term $F(t-R-\epsilon)$ in (\ref{ta16}) whose contribution to
$J_{\nloc}$
  (denoted as  $J_{\reg}$) reads:
\begin{align} \label{ta18}
J_{\reg }=\int
    \frac{R+\epsilon}{R^2\,\sqrt{\epsilon(2R+\epsilon)}} \,
    \frac{\theta(r'-t+R)\,\theta (t-R)}{r'} \;   d^3 \mathbf{r}'\,.
 \end{align}
This differs from the local part (\ref{ta9})  by
$\mathcal{O}\kn(\epsilon^{1/2})$ and thereby the divergent parts
indeed mutually cancel indeed:
\begin{align}
\lim_{\epsilon \to +0}\left( J_{\loc}+J_{\reg }\right)=0.
 \end{align}
\subsection{Finite part}
The remaining part of Eq.~(\ref{ta16}) does not depend on $\epsilon$ and
therefore it is regular in the limit $\epsilon\to 0$:
\begin{align}\label{ta19}
J =\int
   \left[  \left(\vp F(r')-F(0)\right)\, \theta(t-R-r') -F(0) \,
   \theta(r'-t+R)\,\theta (t-R)\right]\,  \frac{d^{\kn 3} \mathbf{r}'}{r'}\,.
 \end{align}
Integration over $\mathbf{r}'$ is performed in spherical coordinates
with the polar angle $\vartheta$  between $\mathbf{r}$ and
$\mathbf{r}'$ and the cyclic azimuthal angle: $d^{\kn 3}
\mathbf{r}'=2\pi\,{r'}^2 \sin{\vartheta}\, dr' d\vartheta$. First we
integrate  over $r'$. In order to reveal   the limits  of
integration, one has to resolve the inequalities $t-R>0$ and
$t-R-r'<>0$  with respect to $r'$. The support of $\theta (t-R)$ is
determined by the solution of the equation $t=R$,
\begin{align} \label{tert}
{r'}^2-2\, r\smhsp r'\cos \vartheta+r^2+z^2-t^2=0
 \end{align}
 with respect to $r'$. The discriminant of this quadratic equation with respect to $r'$
  is
\begin{align}   \label{Ddefi}
  D \equiv r^2 \cos^2 \nhsp
 \vartheta+t^2-r^2-z^2.
  \end{align}
Denoting
\begin{align} \label{ta19a}
\Delta \equiv t^2-r^2-z^2,
 \end{align}
we are led to the description of the domains of integration over
$r',\,\vartheta$ according to different signs of $\Delta$ and $D$. This
analysis is relegated to the Appendix. After careful specification of
the corresponding limits in the double integrals, one can perform both
integrations  in terms of antiderivatives. The final result reads
\begin{align} \label{tb8b}
J=\frac{2\pi\,\theta(t)}{r}
&\biggl[z\,\theta(\Delta)\,\biggl(\arctan\frac{ z\hsp
\sqrt{\Delta}}{t^2\!-\!tr\!-\!z^2} -\arctan\frac{ z\hsp
\sqrt{\Delta}}{t^2\!+\!tr\!-\!z^2}+\pi\,\sgn(z)\,
\theta(tr+z^2-t^2)- \frac{2\,t}{z}\,\arctan\frac{r}{
\sqrt{\Delta}}\biggr) - \nn
\\ &  -\pi\,\theta(-\Delta)\,\theta(t^2-z^2)\,(t-|z|)\biggr]\equiv J_{+\Delta}+J_{-\Delta}\,,
 \end{align}
According to this formula the retarded solution contains  several
sectors divided by the hypersurfaces \mbox{$\Delta=0$} and
\mbox{$z=\pm t$} splitting the spacetime into three regions, as
shown on Fig.\ref{fig1}.
%%%%%%%%%%%%%%%%%%%%%%%%%%%%%%%%%%%%%%%%%%%%%%%%%%%%%%%%%%%%%%%%%%%%%%%%%%%%%%

 \begin{figure}
 \begin{center}
\includegraphics[width=11cm]{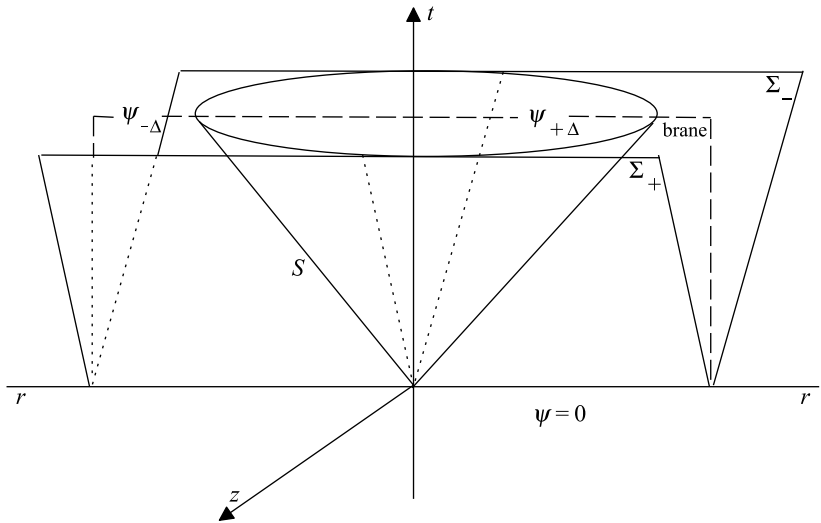}  \caption{Spacetime diagram
of the spherical (the light cone) and the plane-wave fronts in the
bulk. Inside the light cone $S$ defined by $t^2-r^2-z^2=0$, the
solution is $ \psi_{+\Delta}$; outside the light-cone between the
$\Sigma_\pm$ (defined by $t=\pm z$) the solution is $
\psi_{-\Delta}$; outside the interior of two planes and at $t<0$ the
solution is zero.  The function $ \psi$ is discontinuous at the
light cone, but  continuous on the plane fronts.} \label{fig1}
 \end{center}
\end{figure}

%%%%%%%%%%%%%%%%%%%%%%%%%%%%%%%%%%%%%%%%%%%%%%%%%%%%%%%%%%%%%%%%%%%%%%%%%%%%%%%%%%%
In the sector \mbox{$t<|z|$}, $J$ is zero. For $z^2<t^2<r^2+z^2$ it is
determined by the second line of Eq.~(\ref{tb8b}), while for
$t^2>r^2+z^2$ it is determined by the first line. It is easy to check that $J$ is
continuous everywhere at $t>0$:
\begin{itemize}
    \item[i)] The jump of $\pi\, \theta(tr+z^2-t^2)$ in the region \mbox{$\Delta>0$}
    is compensated by the negative
    jump of the  function $\ds \arctan\frac{ z\hsp
\sqrt{\Delta}}{t^2\!-\!tr\!-\!z^2}$.
    \item[ii)] For $t\to|z|$, $J_{-\Delta}$ tends to zero, so $J$
    is continuous on the
    hypersurface $t^2=z^2$.
    \item[iii)] For $t\to \sqrt{r^2+z^2}$, $J_{-\Delta}$ behaves as
    $$\lim_{\Delta\to-0}J_{-\Delta}=-2\pi^2\,\theta(t)\,\frac{t-|z|}{r} \,,$$
     while  in $J_{+\Delta}$ the first two $\arctan$ terms in Eq.~(\ref{tb8b})
     vanish, and the third one
     has the limit $\arctan(r/
\sqrt{\Delta}) \to +\pi/2$. Thus
    $$\lim_{\Delta\to+0}J_{+\Delta}=\frac{2\pi^2\,\theta(t)}{r}\,z  \left[\sgn(z)\,
\theta(r(t-r))-
\frac{t}{z}\right]=\frac{2\pi^2\,\theta(t)}{r}(|z|-t)\,,$$
compensating the above.
\end{itemize}
We will also see that, when the derivatives over $z$ are computed,
there are no delta functions with supports on these boundaries.
\subsection{Physical properties of the solution}
 Differentiating Eq.~(\ref{tb8b}) over $z$ with the help of Eq.~(\ref{zz1}), the
 solution splits as $\psi=\psi_{-\Delta}+\psi_{+\Delta}
 $, where
  \begin{align} \label{tb8}
&\psi_{-\Delta}=  \frac{ \alpha}{ 4 r}
\,\theta(t)\,\theta(t^2-z^2)\,
\theta(-\Delta)\, \sgn z \\
 \label{tb9}
&\psi_{+\Delta}=  \frac{ \alpha}{ 4\pi} \frac{
\theta(t)\,\theta(\Delta)}{ r}\,\biggl[\arctan\frac{z
\sqrt{\Delta}}{t^2-tr-z^2} -\arctan\frac{z
\sqrt{\Delta}}{t^2+tr-z^2}+\pi\,\sgn(z)
\,\theta(-t^2+tr+z^2)\biggr]\,,
 \end{align}
respectively. One can observe  the following.
\begin{itemize}
    \item[i)] The solution contains different regions divided by two characteristic
   four-dimensional hypersurfaces: (i) the light cone $S$, $
   t^2-r^2-z^2=0$, at which the solution is continuous
   and (ii) two planes $\Sigma_\pm\,, t \pm z =0 $,
   at which the solution has jumps. $S$ and $\Sigma_\pm$ touch
   along the lines $z=\pm t, r=0$ (Fig.\,\ref{fig1}).
   Three terms in $\psi_{+\Delta}$ defined inside the cone
   $t^2<r^2+z^2 $ together have no discontinuities on the hypersurface $
   t^2-tr-z^2=0$.

     \item[ii)]  The solution has  support only in the bulk, its restriction
     on the domain wall vanishes.
    In other words, the spherical branon shock wave
    localized  on the
    brane, generates the   \textit{bulk}
    spherical shock wave and two  \textit{planar} shock waves.
All of them start  at the moment of perforation
     $t=0$, propagate in the positive and negative $z$ directions
      and  for $f>0$  carry the positively and
     negatively defined scalar fields, respectively.
     The structure of the characteristic spatial surfaces at fixed $t>0$
     is presented on Fig.\,\ref{fig2}.

     \item[iii)] At a given point $r$ in the bulk one first sees a planar shock
     wave $\Sigma_\pm$ arriving at $t=|z|$, i.e. the jump of $\psi$  form zero to the value
     $\pm  \alpha/{4r}$ for pozitive/negative $z$, respectively
     (the whole  solution is odd in $z$). These values
     are memorized until the arrival of the spherical wave $S$, when
     they start to gradually decay to zero (Fig.\,\ref{t-dep}).
     In other words,  the perforation of the brane
    leaves the  \textit{temporary memory }in the bulk.\kn\footnote{This feature is
    interesting to compare
     with various memory effects in asymptotics of the solutions to d'Alembert
     equations with light-like sources discussed recently
     \cite{Tolish:2014bka,Tolish:2014oda,Bieri:2013ada,Strominger:2014pwa,
     Winicour:2014ska,Bieri:2015yia}.}
The front of the spherical shock wave $S$
     has no discontinuity. For fixed $r$ and
     $z$ after the the passage of the wave front, the solution relaxes with
     time as
     $$ \psi_{+\Delta} \sim   \frac{
   \alpha}{2\pi} \frac{
z}{ t^2 }\,,  $$ which does not depend upon $r$. Therefore for large
enough  $t \gg \sqrt{r^2+z^2}$ the field on the spatial plane
$z={\rm const}$ is \textit{constant} at fixed moment $t$. The plot is
presented on Fig.\,\ref{t-dep}.
\item[iv)] The planar shock waves have the
 structure $ \psi \sim \theta(t\pm|z|)/r$ which differs from the
 Aichelburg-Sexl metric \cite{Aichelburg,GaMeS1} $h_{MN} \sim c_{M} c_{N} \, \delta(t\mp z)/r
 $, where $c^{M}=(1,0,0,0,\pm 1)$ by the degree of discontinuity:
 the Heaviside function versus the delta function. Our shock waves therefore
are ``softer,''
\end{itemize}
%%%%%%%%%%%%%%%%%%%%%%%%%%%%%%%%%%%%%%%%%%%%%%%%%%%%%%%%%%%%%%%%%%%%%%%%%%%
 \begin{figure}
 \begin{center}
\includegraphics[
width=10cm]{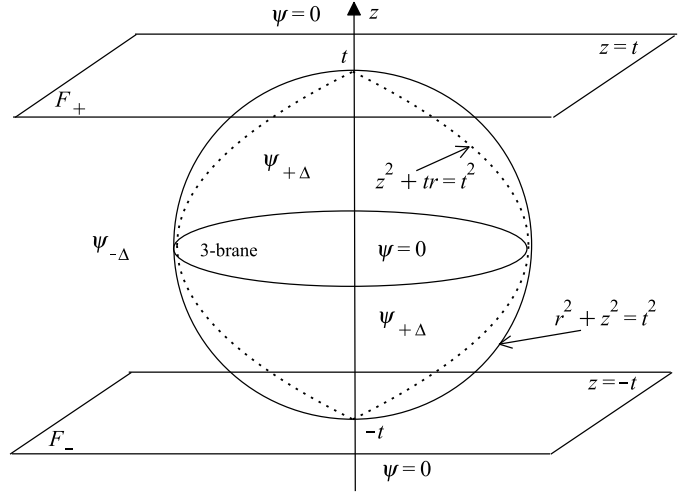}  \caption{ The spatial section of the solution at
fixed time $t>0$. The planes $F_\pm$ are the $t={\rm const}$
sections of $\Sigma_\pm$.
 The surface $z^2+rt=t^2 $ (dotted line) is a solid of
revolution  with the axis $z$ and the parabola with axis "$r$" as a
generatrix  inside the sphere $r^2+z^2=t^2 $.} \label{fig2}
 \end{center}
\end{figure}

 \begin{figure}
 \begin{center}
\includegraphics[
width=9cm]{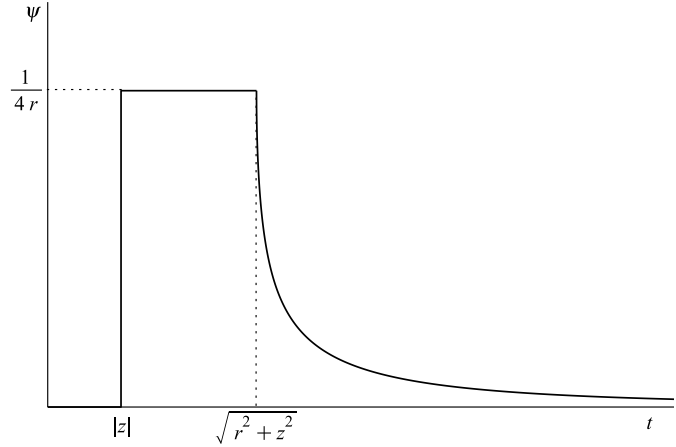}  \caption{ The retarded field $\psi $ as a
function of time at a fixed  observation point ($r>0, z>0$) in the
units $\alpha=1$. The moment $t=|z|$ corresponds to  the passage of the
planar shock wave  leaving as the memory  the the constant value $\psi=
1/{4r}$, which then starts to decay after the passage of the spherical
shock front at the moment $t = \sqrt{r^2+z^2} $.} \label{t-dep}
 \end{center}
\end{figure}
%%%%%%%%%%%%%%%%%%%%%%%%%%%%%%%%%%%%%%
Recall that in this section we have considered the part of the
retarded  scalar field  due to the light-like part of the  source.
For any  relative velocity of the  collision, this part is dominant
in the region of small $k^z$ Eq.\,(\ref{T_br}) in the momentum space. For
ultrarelativistic velocities it is dominant for all momenta.

\setcounter{equation}{0}
\section{Radiation amplitude}
To select  the radiative part of the retarded potential, one has to pass
to the momentum representation of the full current (\ref{source2})
consisting of the local and the stress terms. The first-order
perturbation of the brane stress tensor in the coordinate
representation is given by Eq.\,(\ref{taumn}). Substituting there
the first-order quantity $\un{\!X}^{N} = \Phi\smhsp(\sigma)\,
\delta^{N}_z $, we find
 \begin{align} \label{taumn4}
  \tilde{\rho}(k^M)  =  \fr{f}{4\pi} \!\int \!& \left[\, \vak
\smhsp \bar{h}_{zz}(q) -2\smhsp i\smhsp  k^z \Phi (q) \vp
\right]\delta^{4}(k^{\mu}-q^{\mu})\, d^{\kn 5}q\,,
 \end{align}
where
 \begin{align} \label{NGEQ_mom_f}
 \Phi\hsp(q^{M})=- \frac{i  \pi \vak^2 m }{\gamma}
  \frac{ q^z\smhsp\delta(q^0-v\,q^z)}{q^2 (q_{\mu}q^{\mu}+2i\ep q^0)}  \left( \gamma^2 v^2+\frac{1}{3} \right).
  \end{align}
The integration over $q$ is fulfilled by using delta functions resulting in
 \begin{align}\label{T_br}
  \tilde{\rho}(k^M)=  \frac{\vak^2 f m}{2}\,
 \fr{\gamma v }{\omega^2+k_\bot^2 \gamma^2 v^2}\left(\gamma^2 v^2+\fr{1}{3}\right)\! \left[\frac{k^0\smhsp k^z}{v(k_{\mu}k^{\mu}+2i\ep k^0)}
-1\right],
\end{align}
where $k_\bot^2 \equiv \delta_{ij}k^i k^j$.

The stress part of the second-order current is given by
Eq.\,(\ref{source2}) where only the metric perturbation due to the particle
has to be retained:
\begin{align}\label{source2g}
S(x^M)= \partial_M\!\left(\bar{h}^{MN}
\partial_N-\frac{1}{2}\,\bar{h}\,
\partial^{M}\right)\! \un\varphi\,.
\end{align}
Representing the products of fields as a convolution in the momentum
space, and integrating using delta functions,
 \begin{align}\label{helpp}
\int \left. \frac{\delta(ku-qu) \,\delta^{4}(q^{\mu})}{q^2
(k-q)^2}\,d^5 q \right|_{k^2=0} =\frac{\gamma^3
v^3}{\rule{0cm}{0.9em} (ku)^3(\bar{k}u)}\,, \qquad
\bar{k}^{M}=(k^0,-\mathbf{k},-k_z)\,,
 \end{align}
one obtains
 \begin{align}\label{sf22}
 S(k^M)= \frac{\vak^2 \smhsp f \smhsp m \smhsp \gamma^3 v^3}
 {  (ku) \kn (\bar{k}u) \rule{0cm}{0.9em}}\,.
 \end{align}
We now consider some particular cases.
 \subsection{Ultrarelativistic case}\label{URregime}
Denoting the ratio of the component $k^z$ (orthogonal to the domain
wall) to the total energy of the emitted quantum as $\cos
\chi=k^z/k^0$, we obtain in the  limit  $v \rightarrow 1,\;\gamma
\to \infty$  the local contribution (\ref{T_br}) in the form
 \begin{align}\label{T_br_UR}
 \tilde{\rho}(k^M)=  \frac{\vak^2 f \mathcal{E} }{4 \kn\omega^2
}\,
 \fr{ 1 }{ \cos\chi \cos^2(\chi/2) }\,,
\end{align}
while the stress  term (\ref{sf22}) reads
 \begin{align}\label{sf22v}
 S(k^M)= \frac{\vak^2 f  \mathcal{E}}
 {\om^2 (1-v^2\cos^2\!\chi) \rule{0cm}{0.9em}}\,.
 \end{align}
One  observes the following.
\begin{itemize}
    \item[i)] At $ \chi \approx \pi/2 $ the brane contribution
    dominates and blows up at $ \chi \to  \pi/2 $.
    \item[i)] At $ \chi  <\gamma^{-1} $ in the forward/backward direction the  stress contribution has apparent
    maxima and blows up at $ \chi \to 0$ in the massless limit.
    \item[iii)] At $ \chi \approx \pi $ the brane contribution also blows
    up due to $\cos^2 (\chi/2)$ in the denominator.

\end{itemize}
Considering the last case more closely, we introduce
$\chi'=\pi-\chi$ and find that the local amplitude is negative and
diverges quadratically as $\chi'\to 0$:
 \begin{align}\label{T_br_URa}
 \tilde{\rho}(k^M)=  -\frac{\vak^2 f \mathcal{E} }{4 \kn\omega^2
}\,
 \fr{ 1 }{  \sin^2(\chi'/2) } \simeq  -\frac{\vak^2 f \mathcal{E}
 }{\kn\omega^2  \chi^{\kn \prime}{}^2  }
   \,.
\end{align}
The stress term (\ref{sf22}) is positive and is equal to minus
Eq.\,(\ref{T_br_URa}):
 \begin{align}\label{sf22va}
 S(k^M)= \frac{\vak^2 f  \mathcal{E}}
 {\om^2  \sin^2\!\chi  \rule{0cm}{0.9em}}  \simeq \frac{\vak^2 f  \mathcal{E}}
 {\om^2  \chi^{\kn \prime}{}^2 }\,.
 \end{align}
Thus, as could be expected, radiation is asymmetric with respect to
the particle motion and there is no radiation in the backward
direction.

To get the total radiation power, we note  that (only) the stress amplitude
 is dominant and is beamed around the forward
direction. So to get an estimate of the total power we substitute
Eq.\,(\ref{sf22v}) in
 Eq.\,(\ref{sc_pert10}) and  integrate over angles \textit{for finite
 $\gamma$}.
 The remaining integral over the spectrum diverges both in the
low and high   frequencies requiring the IR and the UV cutoffs:
 \begin{align}\label{sc_pert1011}
E_{\rm rad}= \frac{\vak^4   f^2 m^2  \gamma^2  }{ (2 \pi)^3}
\int\limits_{0}^{\infty} \frac{d\om}{\om}\int\limits_{0}^{\pi/2}
\frac{\sin^2\!\chi \,d\chi}{(1-v^2\cos^2\!\chi)^2}\,\simeq
\frac{\vak^4  f^2  \mathcal{E}^2  \gamma }{8\smhsp(2 \pi)^2} \, \ln
\frac{\om_{\max}}{\om_{\min}}\,.
\end{align}
Note that this expression diverges in the massless limit $\gamma\to 0$,
indicating that the interaction of the brane with massless particles
appeals to quantum theory.

\subsection{Nonrelativistic collision
 }
Now consider the opposite case:  \mbox{$v \ll 1$}. Then the stress
 part (\ref{sf22}) is small, so the total amplitude $j(k)$ is
 given by the brane contribution (\ref{T_br}) only:
 \begin{align}\label{T_br_NR}
j(k)=  \frac{\vak^2 f m}{6\kn \omega }\,  \frac{  k^z}{  k_{\mu}k^{\mu}+2i\ep
 k^0 }\,,
\end{align}
where $\mu=0 \,...\,3$ and the five-dimensional wave vector in null,
so $k_{\mu}k^{\mu}=(k^z)^2,\; \omega^2=(k^z)^2+ k_\perp^2$. This can
be rewritten as
 \begin{align}\label{T_br_NRa}
j(k)=  \frac{\vak^2 f m}{6}\,
 \fr{1}{ k_{z} \sqrt{(k^z)^2+ k_\perp^2}}\,,
\end{align}
so we see that the integration over $k^z$ is dominated in the
infrared, and with logarithmic accuracy  we can set $(k^z)^2+
k_\perp^2 \sim k_\perp^2 $, obtaining
\begin{align}\label{sc_pert10b}
E_{\rm rad} \simeq
 \frac{\vak^4 f^2 m^2}{36\smhsp(2 \pi)^{3}}\,\int
 \frac{dk^z}{ (k_{z})^2} \int dk_\perp \sim  \frac{\vak^4 f^2 m^2}
 {18\smhsp(2 \pi)^{3}}\frac{k_\perp^{\max}}{k^z_{\min}}\,,
\end{align}
where the nature of the cutoff has to be clarified.  Actually, the
divergence of the amplitude in the directions of the emitted quanta
along the brane takes place for any velocity, and its origin is
worth to discussing when going to the coordinate representation.
\subsection{Emission along the brane }\label{revisit}
The general expression for the five-momentum loss under collision in
the coordinate representation is
\begin{align}\label{sc_pert2Bcc}
\Delta P^{M}=\int   \nabla_{N}T^{MN}\, \sqrt{-g}\, d^{\kn 5}
x=\int \varphi^{,M} \rho \,  d^{\kn 5} x\,.
\end{align}
Since the source on the right hand side of  Eq.~(\ref{Phib}) has
support only at $r>t>0$, under the multiplication by $ \psi$ only
the $ \psi_{-\Delta}$part of $ \psi$ will contribute. Computing the
time derivative of $ \psi$ [Eqs.\,(\ref{tb8}) and (\ref{tb9})], we multiply by
a source term (\ref{Phib}), obtaining the estimate
 \begin{align} \label{phi_2_taaa}
 E_{\rm rad}  \sim \frac{
 ( \la f)^2}{ \gamma^6}  \fr{r_{\max}}{z_{\min}}\,,
\end{align}
 where for
consistency and independence of the integration sequence,   the formal
cutoffs for $r$ and $z$ are introduced.

Meanwhile, the validity of our iteration scheme has some
restrictions. The gravitational constant in five dimensions has the
dimension of (length)$^3$. Combining it with the particle energy
$\mathcal{E}$ [dimension of (length)$^{-1}$] and the brane tension
$\mu$ [dimension of (length)$^{-2}$] we have two length parameters: $
l=\vak^2 \mu\,,\; r_S=\vak  \sqrt{\mathcal{E}}\, $, with the first
corresponding to the curvature radius of the bulk in the RS\kn{}II
setup, and  the second
to the gravitational radius of the energy $\mathcal{E}
$.\kn\footnote{Similarly, $1/\vak f$ will be the scalar  analogue of
the curvature length parameter.}  To keep contact with the
RS\kn{}II model we have to consider distances that are small with respect to
$l$, while to justify the linearization of the metric for the
particle we have to consider distances that are large with respect to $r_S$.
So to apply the linearized theory to both objects we have to assume
$ l\gg r_S$, or $\mathcal{E}\ll \vak^2 \mu^2\,. $ With this
motivation we take the maximal cutoff parameters to be
\begin{align}
r_{\max} \sim |z|_{\max} \sim [\vak^2 \mu]^{-1} \,.
 \end{align}

 The minimal values of $r,\;z$ can be estimated from the assumed
convergence of the iterative solution. For the second and the first
terms of  the scalar field, we expect  $ |\de \varphi| \ll  |\un
\varphi| $. Comparing $\de \varphi (\psi)$ [Eq.(\ref{tb8})] with $\un
\varphi $ [Eq.(\ref{ge2A})] for sufficiently small $r$, one finds
\begin{align}\label{rmin}
r_S^2 < r|z|\,.
 \end{align}
On the other hand, we need $ r>r_S$ in order to use the linearized
gravity and the concept of a point-like particle. Combining
these arguments, we conclude that for
\begin{align}\label{rmin1}
 r_{\min}
\sim  z_{\min}\sim r_S\,,
 \end{align}
the perturbation theory converges.

The momentum-space formula (\ref{sc_pert10b})   matches with
Eq.~(\ref{phi_2_taaa}) if
\begin{align}\label{ML}
k^z_{\min} \sim 1/z_{\max}\,, \qquad k_\perp^{\max} \sim
1/r_{\min}\,,
\end{align}
 assuming that the bulk coordinate $z$ is restricted by the same
 curvature effect as the brane coordinate $r$.

With these cutoffs the total radiation loss becomes
\begin{align}\label{sc_pert10d}
E_{\rm rad}
 \sim  \frac{\lambda^2  f^2  }{ \gamma^6
 }\frac{1}{  \vak^3\mu \sqrt{\mathcal{E}}} \sim \frac{\vak  \mathcal{E}^{3/2} f^2  }{   \mu
 }\,,
\end{align}
with the last estimate being valid in the massless-particle limit.

Finally, from the above restrictions  we obtain the frequency
cutoffs:
\begin{align}\label{ML1}
\omega_{\min} \sim  \vak^2\mu\,, \qquad  \omega_{\max} \sim 1/r_{S}
\,.
\end{align}

\subsection{Radiation normal to the brane} Now we come back
 to the massless limit by taking into account  the existence of
$r_{\min}$ preventing the angle $\chi$ from approaching zero:
$$\chi_{\min} \sim \arctan\frac{r_{\min}}{z_{\max}} \simeq
\frac{r_{\min}}{z_{\max}} \ll 1.
$$
In the relativistic case we thus have to integrate the angular
distribution  (\ref{sc_pert1011}) from  $\chi_{\min} $ to $\pi/2$.
Since the integrand is beamed  inside the cone $ 0< \chi \lesssim
1/\gamma $, the final result depends upon the relation between $
1/\gamma$ and $\chi_{\min} $. Namely, by expanding the numerator and
the denominator in Eq.~(\ref{sc_pert1011}) in $ \chi\ll 1$:
$$1-v^2\cos^2\!\chi\simeq \chi^2+\gamma^{-2} \,, $$
we are led to consider two cases:
\medskip

\textbf{Case A: $\chi_{\min}> 1/\gamma$  .} Then,
$1-v^2\cos^2\!\chi\simeq \chi^2 \,, $ so
 \begin{align}\label{tyu1}
E_{\rm rad}= \frac{\vak^4   f^2 m^2  \gamma^2  }{ (2 \pi)^3} \ln
\frac{\om_{\max}}{\om_{\min}}\int\limits_{ \chi_{\min} }^{\pi/2}
\frac{\sin^2\!\chi \,d\chi}{(1-v^2\cos^2\!\chi)^2}\,\simeq
 \vak^4  f^2  \mathcal{E}^2  \,  \frac{r_{\max}}{r_{\min}}  \ln
\frac{r_{\max}}{r_{\min}} \,.
\end{align}
The radiation efficiency $ \epsilon=E_{\rm rad}/\mathcal{E}$ then
reads
 \begin{align}\label{tyu2}
 \epsilon  \sim
 \vak^4  f^2  \mathcal{E}   \,  \frac{r_{\max}}{r_{\min}}  \ln
\frac{r_{\max}}{r_{\min}}\to \vak^2  \mu^2  \mathcal{E}   \,
\frac{r_{\max}}{r_{\min}} \, \ln \frac{r_{\max}}{r_{\min}} \sim
\frac{r_{\min}}{r_{\max}} \ln \frac{r_{\max}}{r_{\min}}<1 \,,
\end{align}
since the function $ \ln x/x$ does not exceed 1 for $ x>1$. Hence
there is no efficiency catastrophe in our model.
\medskip

 \textbf{Case B: $\chi_{\min}< 1/\gamma$  .} Now
$1-v^2\cos^2\!\chi\simeq \gamma^{-2} \,. $ Then the
Eq.~(\ref{sc_pert1011}) holds, but, according to the above
restrictions,
 \begin{align}\label{tyu3}
E_{\rm rad} \sim  \vak^4  f^2  \mathcal{E}^2  \gamma  \, \ln
 \frac{r_{\max}}{r_{\min}}  < \frac{r_{\max}}{r_{\min}}  \ln
\frac{r_{\max}}{r_{\min}}  \,,
\end{align}
so the emitted radiation is smaller than in case A. But now
 radiation is beamed inside the characteristic cone  in the forward
direction  $ \chi\lesssim 1/\gamma$.
\setcounter{equation}{0}
\section{Conclusions}
In this paper we have considered radiation in the
collision of a point particle with an extended
object—-namely, a domain wall—-possessing an internal dynamics.
interaction between them is assumed to be purely
gravitational, while the radiation is scalar and it is generated
solely by the domain wall. The main new feature of this
process is the creation of the shock spherical branon wave,
which propagates freely along the wall with the velocity of
light. This branon constitutes the part of the source of the
scalar radiation. We have carefully calculated the retarded
solution of the bulk  d'Alembert equation with such a
light-like source, revealing a sophisticated structure of the
solution. This solution dominates in the case of the
ultrarelativistic collision.We also computed the full radiation by
taking into account other relevant source terms.

Performing these calculations, we encountered classical
singularities in the solutions of the five-dimensional
d'Alembert equation which are absent in four dimensions.
Namely, the local and the nonlocal parts of the Green's
functions generate two singular parts of the full solution\kn\footnote{  An ambiguity seems to exist
in the literature concerning the local term  of the five-dimensional
Green's function; compare,  e.g.,  with \cite[Eq.\,5]{kazin1}).}
which require regularization. We have then proved that the
sum of these singular parts remains finite when the
regularization is removed. A similar picture holds in higher
odd spacetime dimensions, so our regularity proof may
help in other dimensions too.

The retarded potential of the branon contains two shock plane waves
$\Sigma_\pm $ which have theta-like behavior on the front, differing
from the well-known Aichelburg-Sexl solution~\cite{Aichelburg}
which has a $\delta$ singularity there. Correspondingly, our plane
waves have no point-like sources located at the centers of the
  wave fronts.  The theta-like nature of the planar waves means that after
  the passage of the wave front  the field remains constant, imitating
  the memory effects~\cite{Tolish:2014bka,Tolish:2014oda}.
After the subsequent   arrival of the spherical wave, this value
starts to decrease and asymptotically disappears.
 We realize, however, that this memory is basically due to the nonlocal nature
 of the retarded Green's function in five dimensions.

Finally, we would like to draw attention to one peculiarity of the
radiation amplitude: it exhibits the \textit{backward} destructive
interference  between the local and nonlocal contributions,
contrary to the \textit{forward} destructive interference of the
particle-particle bremsstrahlung \cite{GKST-3, GKST-PLB} and
synchrotron radiation \cite{Khrip, SR}. In an attempt to explain the
difference, one can observe that i) the brane gravity is repulsive, and
ii) the source of radiation is the brane, which moves in the
particle rest frame in the backward direction.

We expect that basic features of the scalar branestrahlung will
survive in the case of the true gravitational radiation.

 \bigskip
\textbf{Acknowledgments.}
This work was supported by the RFBR Grant No.14-02-
01092. The work of P.S. was supported in part by the
European Union's Seventh Framework Programme under
the EU program Thales MIS 375734 and the FP7-REGPOT-
2012-2013-1 No.316165 and was also co-financed by the
European Union (European Social Fund, ESF) and Greek
national funds through the Operational Program Education
and Lifelong Learning of the National Strategic Reference
Framework (NSRF) under the ARISTEIA II Action. Finally,
P.S. is grateful to the noncommercial Dynasty foundation
(Russian Federation) for financial support and to the Crete
 Center for Theoretical Physics at the University of Crete for
hospitality at some stages of this work.
\renewcommand{\theequation}{A.\arabic{equation}}
\setcounter{equation}{0}
\section*{APPENDIX A: \MakeUppercase{Derivation of the of the retarded solution (\ref{ta19})}}

Here we give details of  the calculation of the retarded  field
generated by the light-like branon source (\ref{ta19}).

\medskip
\textbf{Integration domains:} The limits of integration in
$r',\vartheta$ are restricted by several Heaviside functions  in the
integrand depending on the signs of the parameters $\Delta,\;D$
defined in Eqs.~(\ref{Ddefi}) and (\ref{ta19a}).

\smallskip

 \textbf{$\boldsymbol{D>0}$.} If
$\Delta>0$ then $t>r,\;D>0$, and the roots of Eq.~(\ref{tert})
\begin{align} \label{ta20}
r'_{\pm}=r\cos\vartheta  \pm \sqrt{r^2 \cos^2 \nhsp \vartheta+
\Delta}
 \end{align}
(positive and negative respectively) exist for all angles
$\vartheta$. For fixed $t,r,z$ and $\vartheta$, the quadratic form
$R^2-t^2$ as a function of $r'$ goes to $+\infty$ if $r' \to \pm
\infty$, and hence the support of $\theta (t-R)$ lies \textit{between }
the roots. Thus $\theta (t-R)\, \theta(\Delta)=\Theta_{r'}(r'_{-},
r'_{+})\,\theta(\Delta)$. Finally, since $r'_{-}<0$, by assuming $r'>0$
one obtains  $\theta (t-R)\,
 \theta(\Delta) \, \theta(r')=\Theta_{r'}(0, r'_{+})\,\theta(\Delta)$.

The next restricting inequality for the second term of Eq.~(\ref{ta19})
is $R>t-r'$. Being squared, this
 is equivalent to $2\,r'(t-r\cos\vartheta)>\Delta$. If $\Delta>0$ then $t>r>r\cos\vartheta$, so
\begin{align} \label{ta21}
r'> \frac{1}{2} \,\frac{\Delta}{t-r\cos\vartheta} \equiv
\underline{r}'\,,\qquad  \underline{r}'>0\,, \qquad  \theta(r'-t+R)
=\theta(r'-\underline{r}')\,.
 \end{align}
Substituting  \mbox{$r'=\underline{r}'$} into the form
\mbox{$R^2-t^2$} makes it negative. Indeed, being a root of the
equation $r'-t+R(r')=0$, $\underline{r}'$
  satisfies $\underline{r}'-t+ \underline{R} =0$, where $\underline{R}\equiv \ds  R(r'){
\vp}\bigr|_{r'=\underline{r}'}$. Thus $ \underline{R}^2-t^2
=(\underline{R}-t)(\underline{R}+t)=-\underline{r}'(\underline{R}+t)<0$
by virtue of the positiveness of $\underline{r}'$, $t$ and $R$. It
follows that $r'_{-}<\underline{r}'<r'_{+}$. To, summarize: if
$t^2-r^2-z^2>0$, then
\begin{align} \label{ta22}
r'_{-}<0<\underline{r}'<r'_{+} \qquad \theta(r'-t+R)\,\theta (t-R)\,
 \theta(\Delta) \, \theta(r')=\Theta_{r'}(\underline{r}', r'_{+})\,\theta(\Delta)\,.
 \end{align}
Next consider the first term in  Eq.~(\ref{ta19}): according to
Eq.(\ref{ta21}) one obtains:
\begin{align} \label{ta23}
\theta(t-r'-R) =\theta(\underline{r}'-r') \qquad  \theta(t-r'-R)\,
 \theta(\Delta)\, \theta(r') =\Theta_{r'}(0,\underline{r}')\,\theta(\Delta)\,.
 \end{align}
 Thereby making use of $\Theta_{r'}(0,\underline{r}')+
\Theta_{r'}(\underline{r}',r'_{+})=\Theta_{r'}(0, r'_{+})$,  one concludes that the $\theta(\Delta)$ contribution becomes
 \begin{align} \label{ta24}
J_{+\Delta} =2\pi \,\theta(\Delta) \int
   \left[  \vp F(r') \,\Theta_{r'}(0,\underline{r}')-F(0) \,\Theta_{r'}(0, r'_{+})\right]\,  \,{r'}\, dr'\, \sin{\vartheta}\, d\vartheta \,.
 \end{align}

\smallskip

 \textbf{\mbox{$\boldsymbol{\Delta<0,\,D>0}$.}}
This  happens if $
 r^2> t^2-z^2> r^2\,\sin^2 \vartheta\,.
$ Here the equation \mbox{$t=R$} also has two real roots given by
Eq.~(\ref{ta20}) which have the same sign as  $\cos \theta$. Thus if
\mbox{$\cos \theta>0$}, then $\vartheta<\arcsin(\sqrt{t^2-z^2}/r)$,
and $0 <r'_{-} <r'_{+}$; hence
\begin{align} \label{ta26a}
\theta(r')\,\theta(t-R) =\theta(\cos \vartheta)\,\Theta_{r'}(r'_{-}, r'_{+})  \,.
 \end{align}
In the case $\cos \vartheta<0$ one has $r'_{-} <r'_{+}<0$, so
restoring $\theta(r')$, the indicator $ \theta(r')\,\theta(t-R) =0
\,, $ so that the second term in integrand of Eq.~(\ref{ta19}) vanishes.

Now consider the support of $\theta(t-r'-R)$. The limiting value,
$\underline{r}'$ is still determined by Eq.~(\ref{ta21}), but the sign
depends upon the sign of $t-r\cos \vartheta$. If  $t>r\cos
\vartheta$, then $\underline{r}'<0$ by virtue of $\Delta<0$. Here
$\theta(t-r'-R) =\theta(\underline{r}'-r')$ but
\begin{align} \label{ta28}
\theta(r')\,\theta(t-r'-R) =0  \,
 \end{align}
since $\underline{r}'$ is negative. Thus the first term in the
integrand of Eq.~(\ref{ta19})  vanishes, while the integration range for
the second one is determined by $\theta(t-R)$ and reads $r'_{-}< r'
<r'_{+}$.

If  \mbox{$t<r\cos \theta$}, then $\underline{r}'$ [given as before
by the Eq.~(\ref{ta21}] becomes positive, but   this root is
\textit{false} and represents an artifact of the inequality squaring
in Eq.~(\ref{ta21}). Indeed, the formal resolution of $(t-r')^2=R^2$
 gives $r'>\underline{r}'=(r^2+z^2-t^2)/[2\,(r\cos\vartheta-t)]$ and $\underline{r}'>t$
 by virtue of the inequality chain
  $t^2+r^2+z^2 \geqslant t^2+r^2 \geqslant 2\,rt > 2\,rt\cos\vartheta
  $, which is equivalent to $r^2+z^2-t^2>2\,t(r\cos\vartheta-t)$. Meanwhile, $t>r'+R>r'$,
in contradiction with the latter. In what follows, the property
(\ref{ta28}) is valid for all cases of the sign of
$(t-r\cos\vartheta)$.

To conclude: in all cases with ${\Delta<0,\,D>0}$ the first term in
Eq.~(\ref{ta19})  vanishes, while the second is to be integrated over
$r'$ from $r'_{-}$ to $r'_{+}$ with an additional angular
restriction $0\leqslant \vartheta\leqslant \pi/2 $\,.

\smallskip

\textbf{$\boldsymbol{D<0}$.} This restriction is equivalent to $
 t^2-z^2< r^2\,\sin^2 \vartheta
$  and implies \mbox{$\Delta<0$}. The equation \mbox{$t=R$} has no
real roots in this case, so the second term in the integrand of
Eq.~(\ref{ta19}) vanishes completely. The restrictions on
$\theta(t-r'-R)$ are the same as in the previous case, so
$\theta(r')\,\theta(t-r'-R)=0$. Thus one concludes that both terms
in the integrand of Eq.(\ref{ta19})  have no support, and hence the
contribution $J_{D<0}$ to the total $J$ vanishes. \vspace{1em}
Combining the two cases   \mbox{$D>0$} and \mbox{$D<0$} of
$\Delta<0$, one finds
 \begin{align} \label{ta26}
J_{-\Delta} =-2\pi \,\theta(-\Delta) \int
    F(0) \,\Theta_{r'}(r'_{-}, r'_{+})   \,{r'}\, dr'\, \Theta_{\vartheta}(0,\pi/2) \, \theta(D) \,\sin{\vartheta}\, d\vartheta \,.
 \end{align}

\medskip
\textbf{Integration over $\boldsymbol{r'}$:} We have to integrate
$J_{-\Delta}$, given by Eq.~(\ref{ta26}), and $J_{+\Delta}$ from
Eq.~(\ref{ta24}). First consider $J_{-\Delta} $ : integrating
Eq.~(\ref{ta26})   from $r'_{-}$ to  $r'_{+}$  with help of the table
integral
 \begin{align}\label{ta24chj}
\int \frac{(\alpha x+\beta)\, dx}{(x^2+b^2)
\sqrt{a^2-x^2}}=\frac{1}{\sqrt{a^2+b^2}} \left[\frac{\beta}{b}
\,\arctan \frac{x\,\sqrt{a^2+b^2} }{b\,\sqrt{a^2-x^2}}-
\alpha\,\arctan \frac{\sqrt{a^2-x^2}}{\sqrt{a^2+b^2}}\right]
 \end{align}
one arrives at
 \begin{align} \label{ta24d}
J_{-\Delta}  =-2\pi^2 \,r\,\theta(t)\,\theta(-\Delta) \int
\frac{ \theta(D)\,\cos \vartheta\, \sin{\vartheta}\,
}{\sqrt{{z}^{2}+{r}^{2}
   \sin^{2}\nhsp \vartheta }}\,
  d\vartheta \,,
 \end{align}
with the remaining integral over $\vartheta$.

The contribution $J_{+\Delta}$  consists of two parts: $J_{+\Delta}
\equiv Q_1+Q_2$. The one containing $F(0)$  is of the  type
(\ref{ta24chj}), so by integrating it from 0 to $r'_{+}$ we obtain
\begin{align}
 \label{ta65}
Q_2=-\frac{1}{2}\,\ln\frac{t+\sqrt{\Delta}}{t-\sqrt{\Delta}}-\frac{r\,\cos\vartheta}{\sqrt{{z}^{2}+{r}^{2}
   \sin^{2}\nhsp \vartheta }}\,\left( \arctan\frac{rt\,\cos\vartheta}{ \sqrt{\Delta \, \smash{(}  z^{2}+r^{2}
   \sin^{2}\nhsp \vartheta \smash{)} }}
+  \frac{\pi}{2}\right).
\end{align}
The first contribution in Eq.~(\ref{ta24}) coming from the $F(r')$ term
contains the integrals of type
 \begin{align}\label{ta24chl}
\int \frac{(\alpha x^2+\beta x+\gamma)\, dx}{(x^2+b^2)
\sqrt{c-x}}\,.
 \end{align}
A routine  calculation gives
 \begin{align}
 %\label{ta24ch2}
Q_1=\frac{1}{2}\,\ln\frac{t+\sqrt{\Delta}}{t-\sqrt{\Delta}}-\frac{\sqrt{\Delta}}{t-r\cos\vartheta}+\frac{r\,\cos\vartheta}{\sqrt{{z}^{2}+{r}^{2}
   \sin^{2}\nhsp \vartheta }}\,\left( \arctan\frac{t-r\cos\vartheta+\sqrt{\Delta}}{ \sqrt{
{{z}^{2}+{r}^{2}
   \sin^{2}\nhsp \vartheta }}} - \arctan\frac{t-r\cos\vartheta-\sqrt{\Delta}}{ \sqrt{ {{z}^{2}+{r}^{2}
   \sin^{2}\nhsp \vartheta }}}
\right).\nn
 \end{align}
Combining it with  Eq.~(\ref{ta65}) and using the identity
 \begin{align}
 \label{ggggg}
\arctan x - \arctan y= \arctan  \frac{x-y}{1+xy}+\pi \hsp
\theta(-1-xy)\,\sgn x
 \end{align}
we get
 \begin{align}
 \label{ta24ch3}
J_{+\Delta} =& 2\pi \,\theta(\Delta)  \int\Biggl[
-\frac{\sqrt{\Delta}}{t-r\cos\vartheta}+\frac{r\,\cos\vartheta}{\sqrt{{z}^{2}+{r}^{2}
   \sin^{2}\nhsp \vartheta }}  \, \times   \\ & \times    \Biggl( \arctan\frac{ \sqrt{\Delta} \sqrt{z^2+r^2
   \sin^2\nhsp \vartheta }}{r^2 +z^2 -rt\hsp\cos\vartheta}
   +\frac{\pi}{2}\:
   \sgn(rt\hsp\cos\vartheta-z^2-r^2)
-\arctan\frac{rt\,\cos\vartheta}{\sqrt{\Delta \, \smash{(}
z^{2}+r^{2}
   \sin^{2}\nhsp \vartheta \smash{)} }}
\Biggr)\Biggr]\sin\vartheta \,d\vartheta\,.  \nn
 \end{align}

\medskip

\textbf{Integration over $\boldsymbol{\vartheta}$:} The condition
$D>0$ is equivalent to $t^2-z^2>r^2\sin^2\nhsp \vartheta>0$ in
addition to $\theta(-\Delta)$. Thus one  integrates over $\vartheta$
from zero to $\arcsin( \sqrt{t^2-z^2}/r)$. A thorough analysis of the
characteristic domains gives in this case a simple overall condition
$t>|z|$ :
 \begin{align} \label{tb6}
\theta(-\Delta)\,\theta(D)\,
=\theta(-\Delta)\,\theta(t^2-z^2)\,\Theta_{\vartheta}\nhsp\biggl(0,
\arcsin\frac{ \sqrt{t^2-z^2}}{r}\biggr).
 \end{align}
Using this in the angular integral in Eq.~(\ref{ta24d})   we obtain
 \begin{align} \label{tb7}
J_{-\Delta}=-2\pi^2\,\theta(t)\,\theta(-\Delta)\,\theta(t^2-z^2)\,
\frac{t-|z|}{r}\,.
 \end{align}
For the region $\Delta>0$  one has to integrate  Eq.~(\ref{ta24ch3}).
The first term in the brackets  yields
 \begin{align} \label{zz0}
 -\int\limits_0^\pi
 \frac{\sqrt{\Delta}\, \sin\vartheta}{t-r\cos\vartheta}
 \,d\vartheta=\frac{\sqrt{\Delta}}{r}\, \ln\frac{t-r}{t+r}\,.
 \end{align}
The second one is integrated by parts, taking into account the jump
of the arctangent:
 \begin{align} \label{zz1}
 \frac{d}{dx}\,\arctan \frac{1}{x}=-\frac{1}{1+x^2}+\pi
 \,\delta(x)\,.
 \end{align}
One obtains
$$
J_{+\Delta}=\frac{2\pi\,z\,\theta(t)\,\theta(\Delta)}{r}
\biggl[\arctan\frac{ z\hsp \sqrt{\Delta}}{t^2\!-\!tr\!-\!z^2}
-\arctan\frac{ z\hsp
\sqrt{\Delta}}{t^2\!+\!tr\!-\!z^2}+\pi\,\sgn(z)\,
\theta(tr+z^2-t^2)- \frac{2\,t}{z}\,\arctan\frac{r}{
\sqrt{\Delta}}\biggr].
$$
 Combining this with Eq.~(\ref{tb7}), we find the result (\ref{tb8b}) presented in the main text.

\end{document}